\documentclass[structabstract]{aa}
\usepackage{graphicx}
\usepackage{natbib}
\newcommand{\chir}{\ensuremath{\chi_\nu^{\,2}}}                   % Reduced chi-squared symbol

\usepackage{amsmath,amssymb,graphicx}
\begin{document}

\title{Simultaneous follow-up of planetary transits: revised physical properties
for the planetary systems HAT-P-16 and WASP-21}

%   \subtitle{}
\titlerunning{HAT-P-16\,b \& WASP-21\,b}

   \author{
          S. Ciceri\inst{1}
          \and
          L. Mancini \inst{1} %\thanks{}
          \and
          J. Southworth\inst{2}
          \and
          N. Nikolov\inst{1,3}
          \and
          V. Bozza\inst{4,5}
          \and
          I. Bruni\inst{6}
          \and \\
          S. Calchi Novati\inst{4,7}
          \and
          G. D'Ago\inst{4}
          \and
          Th. Henning\inst{1}
          }

       \institute{Max Planck Institute for Astronomy, K\"{o}nigstuhl 17, 69117 -- Heidelberg, Germany \\
             \email{ciceri@mpia.de}
         \and
    Astrophysics Group, Keele University, Staffordshire, ST5 5BG, UK
        \and
    Astrophysics Group, University of Exeter, Stocker Road, EX4 4QL, Exeter, UK
        \and
    Department of Physics, University of Salerno, Via Ponte Don Melillo, 84084 -- Fisciano (SA), Italy
        \and
    Istituto Nazionale di Fisica Nucleare, Sezione di Napoli, Napoli, Italy
        \and
    INAF -- Osservatorio Astronomico di Bologna, Via Ranzani 1, 40127 -- Bologna, Italy
        \and
    Istituto Internazionale per gli Alti Studi Scientifici (IIASS), Vietri Sul Mare (SA), Italy
}

%   \date{Received ; Accepted}
% \abstract{}{}{}{}{}
% 5 {} token are mandatory
  \abstract
{By now more than 300 planets transiting their host star have been found, and much effort is being put into measuring the properties of each system. Light curves of planetary transits often contain deviations from a simple transit shape, and it is generally difficult to differentiate between anomalies of astrophysical nature (e.g.\ starspots) and correlated noise due to instrumental or atmospheric effects. Our solution is to observe transit events simultaneously with two telescopes located at different observatories.}
{Using this observational strategy, we look for anomalies in the light curves of two transiting planetary systems and accurately estimate their physical parameters.}
{We present the first photometric follow-up of the transiting planet HAT-P-16\,b, and new photometric observations of WASP-21\,b, obtained simultaneously with two medium-class telescopes located in different countries, using the telescope defocussing technique. We modeled these and other published data in order to estimate the physical parameters of the two planetary systems.}
{The simultaneous observations did not highlight particular features in the light curves, which is consistent with the low activity levels of the two stars. For HAT-P-16, we calculated a new ephemeris and found that the planet is 1.3 $\sigma$ colder and smaller ($R_\mathrm{b}=1.190 \pm 0.037 \; R_\mathrm{Jup}$) than the initial estimates, suggesting the presence of a massive core. Our physical parameters for this system point towards a younger age than previously thought. The results obtained for WASP-21 reveal lower values for the mass and the density of the planet (by 1.0 and 1.4 $\sigma$ respectively) with respect to those found in the discovery paper, in agreement with a subsequent study. We found no evidence of any transit timing variations in either system.}
{}
{}


\maketitle

% Sect. 1
%%%%%%%%%%%%%%%%%%%%%%%%%%%%%%%%%%%%%%%%%%%%%%%%%%%%%%
\section{Introduction}
\label{sec_1}
%%%%%%%%%%%%%%%%%%%%%%%%%%%%%%%%%%%%%%%%%%%%%%%%%%%%%%

Since the discovery of the first planet orbiting a main sequence star \citep{maiorequeloz1995}, more than 900 extrasolar planets have been found using different techniques. It is therefore possible to analyse these planets from a statistical viewpoint, and compare the predictions of theoretical models (e.g.\ \citealp{fortney2007,liu2011,mordasini2012a,mordasini2012b}) to the available data (e.g.\ \citealp{gould2010,mayor2011,howard2012,cassan2012,fressin2013}). Such comparisons are fundamental to confirm or discard different theories of planet formation and evolution.

Whilst it is important to enlarge the number of detected planets, it is also vital to accurately measure the main physical properties of each planetary system used in statistical analysis. In this context, the transiting extrasolar planets (TEPs) are extraordinary sources of information. The particular geometry of these systems enables measurement of the complete set of their physical properties (e.g.\ \citealp{charbonneau2000,henry2000}). From the light curve obtained during transit events, it is possible to measure the inclination of the orbit with respect to the line of sight and the size of the system's components. Combining these results with spectroscopic measurements, we can obtain a precise value of the mass of the planet (rather than just a lower limit as for extrasolar planets detected by radial velocity measurements; \citealp{marcy1998}).

TEPs are also the only extrasolar planets for which we can investigate the atmospheric composition, both with spectroscopy during transit and occultation (e.g.\ \citealp{charbonneau2002,deming2005,knutson2007,swain2008,borucki2009}), and by multi-band photometric observations with the aim of detecting variation of the planet's radius as a function of wavelength \citep{southworth2012al,mancini2013a,mancini2013b,mancini2013c,nikolov2013}. Moreover, the presence of an additional body orbiting the host star can be inferred based on transit timing variation studies (e.g.\ \citealp{holman2010,steffen2013,maciejewski2013}), which require a large sample of accurate mid-transit times.

Anomalies in the light curves of planetary transits can arise from several phenomena affecting the parent stars, such as gravity darkening \citep{barnes2009,szabo2011}, stellar pulsation \citep{cameron2010}, starspots (e.g.\ \citealp{sanchis2011,tregloan2013}) and even the presence of exomoons \citep{kipping2009}. High-quality photometric observations are therefore not only important to accurately determine the physical parameters of TEP systems, but can yield further astrophysical information. However, even if we use the telescope-defocussing method, which allows a much better photometric precision than traditional in-focus photometry \citep[e.g.][]{tregloan2013b}, it is generally a hard task to distinguish transit anomalies due to astrophysical effects from those caused by random or systematic noise attributable to instrumental or atmospheric effects.

One solution is to monitor the same transit event simultaneously from two telescopes located at different sites. If data from both the telescopes contain the same anomaly, we can discard the possibility that it is caused by instrumental effects or effects due to Earth's atmosphere. We successfully implemented this observational strategy to follow up several planetary transits by using the Cassini 1.5\,m telescope at the INAF/Astronomical Observatory of Bologna in Loiano (Italy) and the CA 1.23\,m telescope at the German-Spanish Astronomical Center at Calar Alto (Spain). These two telescopes are sufficiently distant from each other that their observations are completely independent in terms of instrumental effects and atmospheric conditions, but close enough that they can contemporaneously observe the same transit event. Additionally, the new data should provide a better estimation of the photometric parameters. We tested this approach in 2011, when we observed a transit of HAT-P-8 simultaneously with the two telescopes. We did indeed notice an asymmetry into the light curve, and its presence in both datasets confirms the reality of the signal \citep{mancini2013a}. Here we present the results of this simultaneous-observation approach applied to two planetary systems, HAT-P-16 and WASP-21, both of which host a close-in gaseous TEP.

% Sect. 1.1
%%%%%%%%%%%%%%%%%%%%%%%%%%%%%%%%%%%%%%%%%%%%%%%%%%%%%%
\subsection{Case history}
\label{sec_1.1}
%%%%%%%%%%%%%%%%%%%%%%%%%%%%%%%%%%%%%%%%%%%%%%%%%%%%%%

% HAT-P-16

HAT-P-16\,b was detected by \citet{buchhave2010}, who found it to be a be a $4.2\,M_{\mathrm{Jup}}$ hot Jupiter on a slightly eccentric 2.8\,day orbit ($e=0.036$) around a $V=10.8$ mag, F-type star. With a radius of $1.3 \, R_{\mathrm{Jup}}$ the planet is nearly twice as dense as Jupiter. The Rossiter-McLaughlin effect has been detected by \citet{moutou2011} who found a projected spin-orbit angle of $\lambda=-10^{\circ} \pm 16^{\circ}$, which is consistent with a prograde, aligned orbit.

% WASP-21

The WASP-21 system hosts a hot Saturn-like planet with a mass of $0.3 \, M_{\mathrm{Jup}}$ and radius of $1.1 \, R_{\mathrm{Jup}}$ \citep{bouchy2010}. The planet moves on a circular orbit with a period of $\sim 4.32$\,days around a $V=11.58$ mag, G-type star. The parent star has a mass and radius similar to the Sun, but also one of the lowest metallicities known ($[\frac{\mathrm{Fe}}{\mathrm{H}}]= - 0.46$) for a TEP host star. The physical parameters of this system were revised by \citet{barros2011},who found that the WASP-21\,A star is evolving off the main sequence and, depending on the assumptions made in the analysis used, has a lower density than found in the discovery paper. The revised planetary properties pointed to a lower mass and slightly larger radius. The low density implies the planet is coreless and has a H/He composition.

In this work we present simultaneous transit observations of these two planetary systems from two telescopes. In Sect.\,\ref{sec_2} we show the first follow-up observations of HAT-P-16 and new photometric data for WASP-21. We used these new light curves to revise the physical parameters of these two TEP systems. The details of the light curve analysis are described in Sect.\,\ref{sec_3}, whereas our estimations of the physical parameters are reported in Sect.\,\ref{sec_4}. The results of our work are summarized in Sect.\,\ref{sec_5}.

%  Sect. 2

%%%%%%%%%%%%%%%%%%%%%%%%%%%%%%%%%%%%%%%%%%%%%%%%%%%%%%

\section{Observations and data reduction}

\label{sec_2}

%%%%%%%%%%%%%%%%%%%%%%%%%%%%%%%%%%%%%%%%%%%%%%%%%%%%%%

\begin{figure}%

\centering

\includegraphics[width=8.cm]{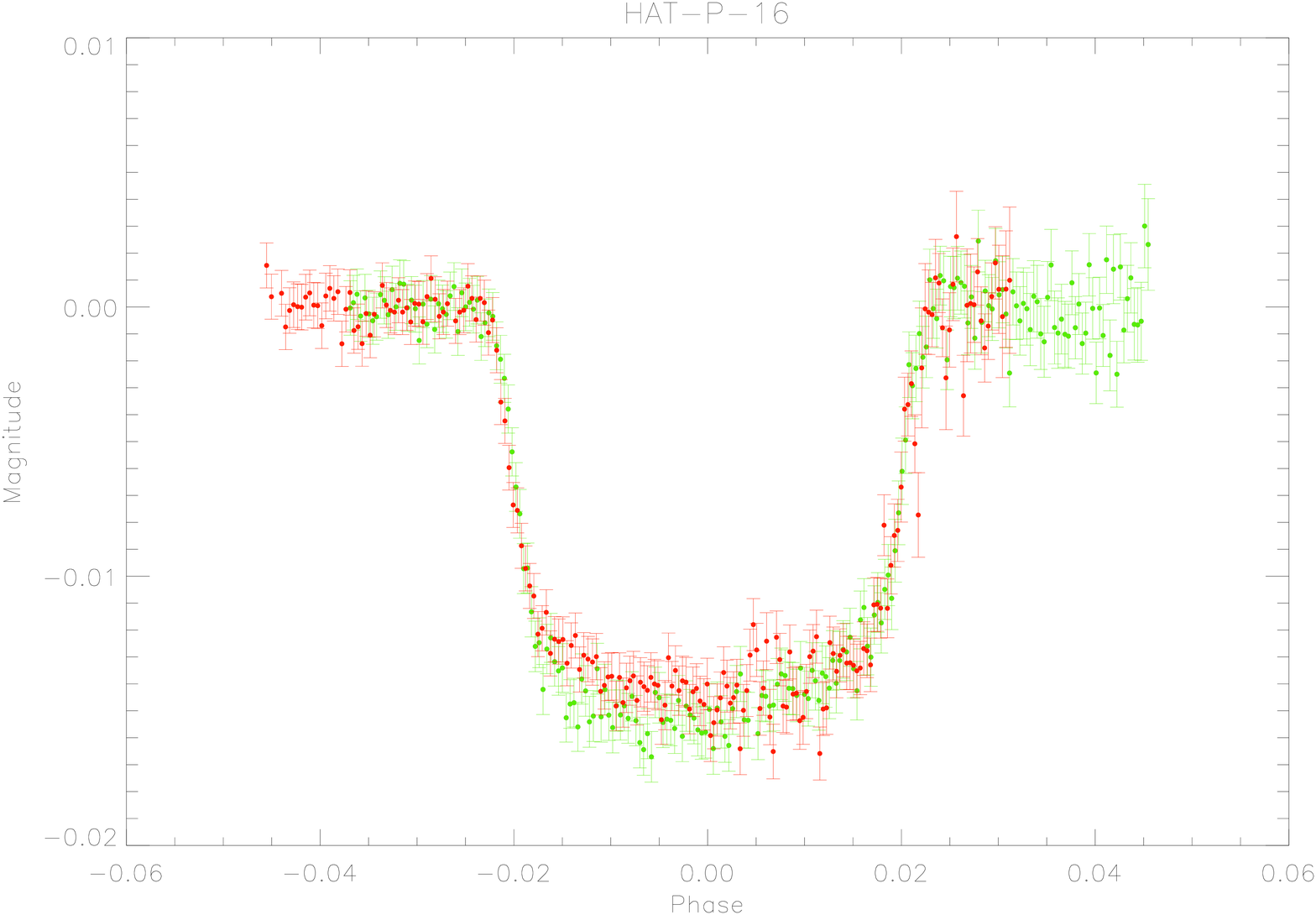}

\caption{Observations of the October 2012 transit of HAT-P-16. The green points show the data from the CA 1.23\,m telescope, and the red points the data from the Cassini telescope.}%

\label{sup_hp16}

\end{figure}

\begin{figure}%

\centering

\includegraphics[width=8.cm]{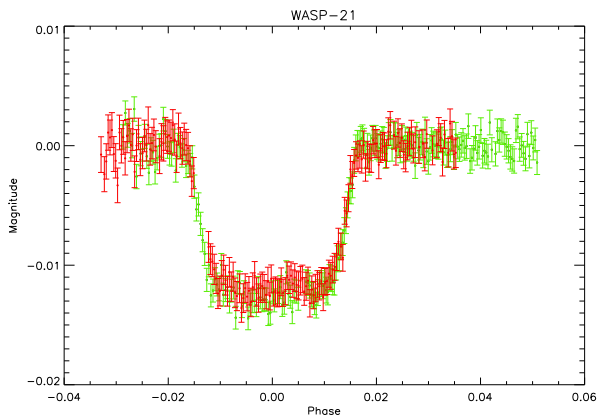}

\caption{Observations of the September 2012 transit of WASP-21. The green points show the data from the CA 1.23\,m telescope, and the red points the data from the Cassini telescope.}%

\label{sup_w21}

\end{figure}

For both planetary systems, we observed one transit event simultaneously with two telescopes (Figs.\ \ref{sup_hp16} and \ref{sup_w21}). These observations were carried out between September and October 2012 from the Loiano and Calar Alto observatories. An additional transit of HAT-P-16 was observed on October $29^{\mathrm{th}}$ 2010 from Loiano during the PLAN microlensing campaign towards M31 \citep{novati2009,novati2010}. Another transit of HAT-P-16 was observed in Calar Alto on August $22^{\mathrm{th}}$ 2011. In total we present six new light curves, five of them being from defocussed 1.2--1.5\,m telescopes (see Table \ref{ObsLog}). With the telescope-defocussing technique we can use larger exposure times ($\sim 50-120$\,s) which allows us to collect many more photons over a large number of pixels, thus reducing the Poisson and scintillation noise and minimizing systematic noise due to flat-fielding errors, seeing variations and image motion \citep{southworth2009al1,southworth2009al2}. This is particularly useful for planetary transits, because the variation in flux of the star due to the planet passing in front of it is small, generally 2\% or less.

% Table 1

\begin{table*}

\centering %

\caption{Observing log for the two TEP systems. $N_{\rm obs}$ is the number of observations, $T_{\rm exp}$ is the exposure time, $T_{\rm obs}$ is the observational cadence, and `Moon illum.' is the fractional illumination of the Moon at the midpoint of the transit.}

\tiny %

\begin{tabular}{lccccccccccc}

\hline %

\hline %

Telescope & Date of   & Start time & End time  &$N_{\rm obs}$ & $T_{\rm exp}$ & $T_{\rm obs}$ & Filter & Airmass & Moon & Aperture   & Scatter \\

        & first obs &    (UT)    &   (UT)    &              & (s)           & (s)           &        &         &illum.& radii (px) & (mmag)  \\

\hline %

\multicolumn{10}{l}{HAT-P-16:} \\

Cassini   & 2010 10 29 & 21:19 & 03:01 & 261 & 50--60   &  70 & Johnson $I$ & 1.00 $\to$ 2.16 & 56\% & 17, 38, 58 & 1.01 \\

CA\,1.23m & 2011 08 22 & 23:08 & 04:36 & 537 & 11--15   &  39 & Cousins $R$ & 1.00 $\to$ 1.33 & 39\% & 15, 40, 60 & 1.85 \\

Cassini   & 2012 10 03 & 23:28 & 04:41 & 197 & 70--80   &  92 & Johnson $I$ & 1.01 $\to$ 2.15 & 86\% & 18, 50, 70 & 0.96 \\

CA\,1.23m & 2012 10 04 & 00:02 & 05:31 & 216 & 70--80   &  90 & Cousins $I$ & 1.00 $\to$ 2.04 & 86\% & 24, 38, 60 & 0.88 \\

\hline %

\multicolumn{10}{l}{WASP-21:} \\

Cassini   & 2012 09 11 & 19:06 & 02:10 & 177 & 120      & 135 & Gunn $i$    & 1.72 $\to$ 1.11 & 18\% & 17, 28, 46 & 0.88 \\

CA\,1.23m & 2012 09 11 & 19:28 & 03:48 & 218 & 100--130 & 135 & Cousins $I$ & 1.63 $\to$ 1.08 & 18\% & 22, 40, 65 & 0.86 \\

\hline

\end{tabular}

\label{ObsLog}%

\end{table*}

% Figure 01

%

% Figure 02

%

%%%%%%%%%%%%%%%%%%%%%%%%%%%%%%%%%%%%%%%%%%%%%%%%%%%%%%

\subsection{1.52\,m Cassini Telescope }

\label{sec_2.1}

%%%%%%%%%%%%%%%%%%%%%%%%%%%%%%%%%%%%%%%%%%%%%%%%%%%%%%

One transit event of WASP-21 and two of HAT-P-16 were observed with the Cassini telescope. This 152\,cm telescope is located at the Loiano Observatory near Bologna (Italy), and was already successfully used to follow up several planetary transits (e.g.\ \citealp{harpsoe2012,  southworth2012al2}). It has an German-type equatorial mount with a Ritchey-Chr\'etien configuration. It is equipped with the BFOSC (Bologna Faint Object Spectrograph \& Camera), whose CCD has $1300 \times 1340$ pixels and a plate scale of $0.58^{\prime \prime}$ pixel$^{-1}$, resulting in a a field of view of $13^{\prime} \times 12.6^{\prime}$. For all observations, the CCD was windowed to decrease the readout time and the telescope was defocussed and autoguided. For the WASP-21 transit we used a Gunn-$i$ filter, while the HAT-P-16 ones were observed through a Johnson-$I$ filter.

The science images were bias subtracted and flat-fielded. Master bias and flat-field images were created combining multiple suitably scaled bias and flat images. The bias and the flat-field frames were collected during the same nights as the observations. In particular the flat-field frames were taken on the sky immediately after sunset. Light curves were extracted using an aperture-photometry routine based on the \texttt{DAOPHOT} photometry package \citep{stetson1987} and IDL's \texttt{astrolib/APER} routine. We tried different values for the circular apertures in order to find the most precise photometry, i.e.\ the light curve with the smallest scatter in the out-of-transit region. We noticed that changes in the aperture size do not have a significant effect on the shape of the light curves but do cause small differences in the scatter of the datapoints. The apertures used for each transit are reported in Table\,\ref{ObsLog}. For fixed aperture size we also tried different numbers of comparison stars to obtain differential photometry. The final comparison stars were chosen according to their brightness and the scatter in the resulting light curve.

%%%%%%%%%%%%%%%%%%%%%%%%%%%%%%%%%%%%%%%%%%%%%%%%%%%%%%

\subsection{1.23\,m Calar Alto Telescope}

\label{sec_2.2}

%%%%%%%%%%%%%%%%%%%%%%%%%%%%%%%%%%%%%%%%%%%%%%%%%%%%%%

We observed two transits of HAT-P-16 and one of WASP-21 with the 1.23\,m telescope at Calar Alto. Mounted in the Cassergrain focus of this telescope is the new DLR-MKIII camera, which has $4000 \times 4000$ pixels, a plate scale of $0.32^{\prime \prime}$ pixel$^{-1}$ and a field of view of $21.5^{\prime} \times 21.5^{\prime}$. This instrumental equipment was already successfully used to observe planetary transits (e.g.\ \citealp{mancini2013a,maciejewski2013}). For the two transits observed in 2012 we used a Cousins-$I$ filter, the telescope was autoguided and defocussed, and the CCD was windowed (Table\,\ref{ObsLog}). For the transit observed on August the $22^{\mathrm{th}}$ 2011 a different camera with a smaller field of view (the SITE\#2b CCD, with a plate scale of 0.51 arcsec per pixel) was used with a Cousins-$R$ filter, and the telescope was autoguided but not defocussed. The observations were reduced in the same way as those from the Cassini Telescope.

\section{Light curve analysis}

\label{sec_3}

%%%%%%%%%%%%%%%%%%%%%%%%%%%%%%%%%%%%%%%%%%%%%%%%%%%%%%

To measure the photometric parameters of the systems, all the light curves were fitted individually following much of the methodology of the \emph{Homogeneous Studies} project (\citealp{southworth2012}; and references within). The fits were performed using the {\sc jktebop} code \citep{southworth2008}, which models the two components of the planetary system as biaxial spheroids. The main parameters fitted by the code are the orbital inclination $i$, the time of transit midpoint $T_{0}$, the sum of the reduced radii (`reduced' radius is the ratio between the true radius of the object and the semi-major axis $a$ of the orbit) $r_{\mathrm{A}}+r_{\mathrm{b}}$, and the ratio of the radii $k = r_{\mathrm{b}}/r_{\mathrm{A}}$.

An important effect to consider when fitting transit light curves is limb darkening (LD). We used the quadratic LD law and obtained theoretical coefficients from interpolation in the tables of \citet{claret2003}. We tried two different strategies: ($i$) fitting the linear LD coefficient and fixing the quadratic coefficient to the theoretical value; ($ii$) fixing both LD coefficients to the theoretical values. From the two analysis we kept the results with greatest internal consistency.

For WASP-21 we adopted a circular orbit, following the findings of \citet{barros2011} and \citet{pont2011}. By contrast, HAT-P-16 has a well-established orbital eccentricity, $e$, and longitude of periastron, $\omega$. We adopted the constraints $e\cos\omega = -0.030 \pm 0.003$ and $e\sin\omega = -0.021 \pm 0.006$ \citep{buchhave2010} to include the effects of an eccentric orbit in the light curve fits.

\begin{figure}%
\centering
\includegraphics[width=8.cm]{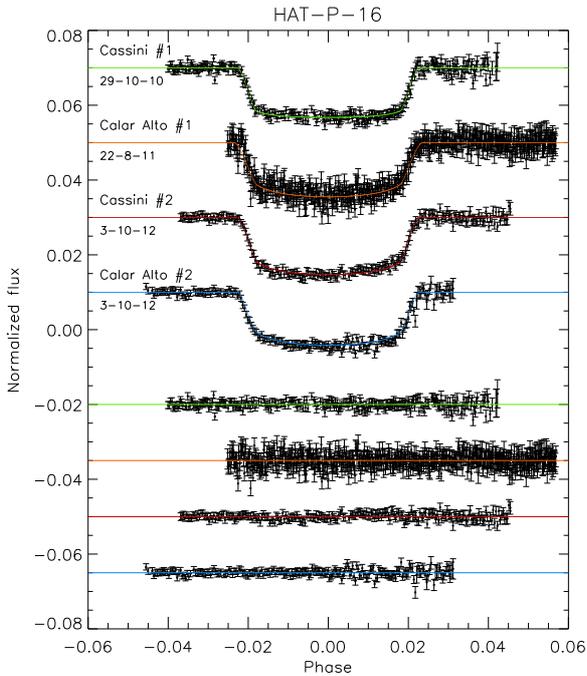}
\caption{Light curves of HAT-P-16 compared with the best {\sc
jktebop} fits. The dates and instruments used for each transit
event are indicated. Residuals from the fits are displayed at the
bottom, in the same order as the top curves.}%
\label{fit_hp16}
\end{figure}

\begin{figure}%
\centering
\includegraphics[width=8.cm]{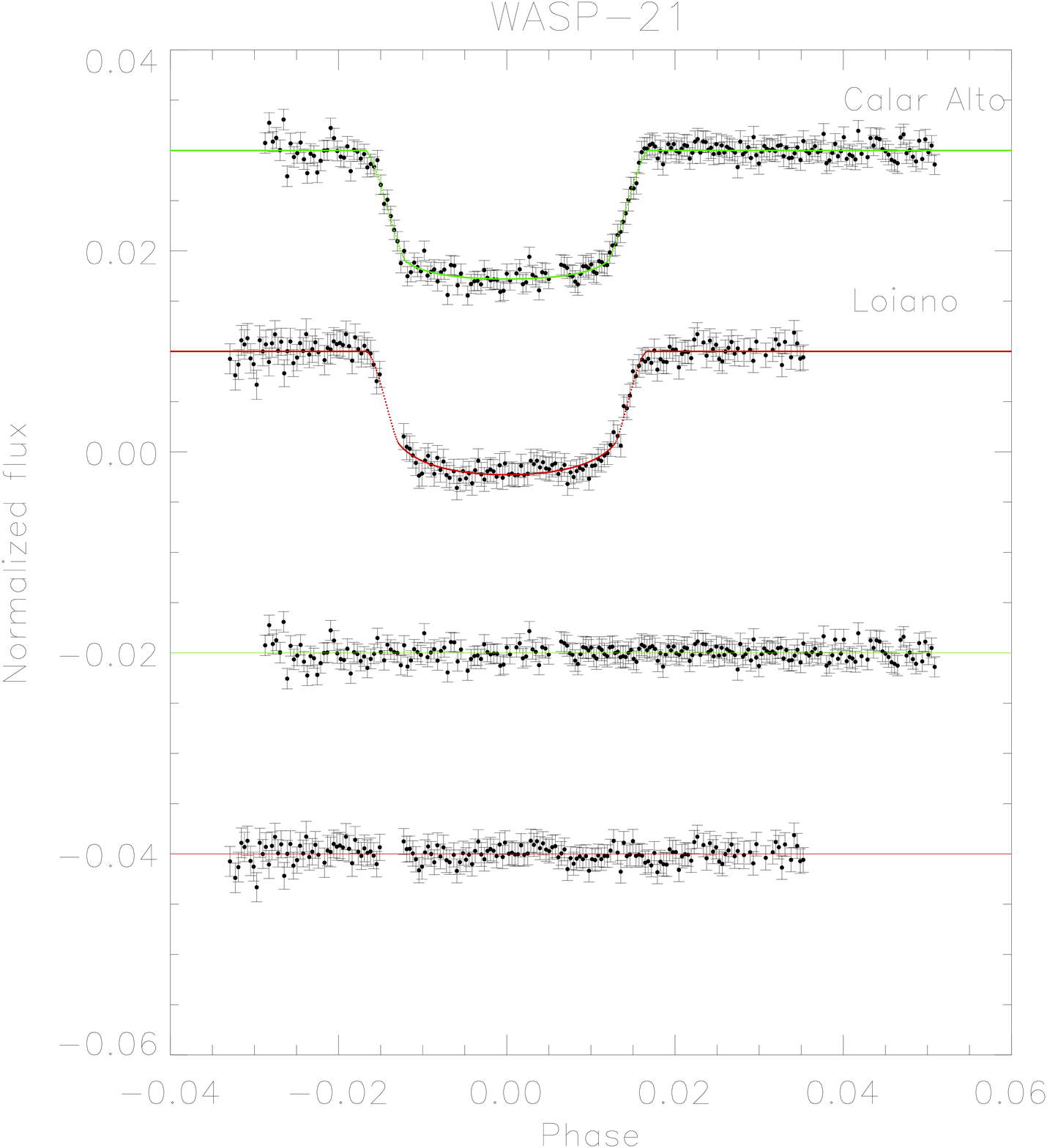}
\caption{Light curves of WASP-21 compared with the best {\sc
jktebop} fits. The instruments used for each transit
event are indicated. Residuals from the fits are displayed at the
bottom, in the same order as the top curves.}%
\label{fitt_w21}
\end{figure}

In order to take into account the red noise and compensate for the underestimated errorbars produced by the {\sc aper} algorithm, we performed a two-step inflation of the errorbars, as used in several published studies (e.g.\ \citealp{gibson2008,winn2008,winn2009,nikolov2012,mancini2013a,mancini2013b}). It consists of running the fitting code once for each light curve and then rescaling the errorbars of each dataset to give a reduced $\chi^2$ of $\chi_{\nu}^{2}=1$. The errorbars are then further inflated through the $\beta$ approach \citep{pont2006,gillon2006,winn2007}. We then ran {\sc jktebop} once more on these error-rescaled datasets, obtaining the final values of the parameters, which are reported in Tables \ref{pho-res_hp16} and \ref{pho-res_w21}. The light curves and best-fitting models are shown in Fig.\,\ref{fit_hp16} for HAT-P-16, and Fig.\,\ref{fitt_w21} for WASP-21. The residuals of each fit are plotted at the bottom of the figures.

% Figure 03
%
%
% Figure 04

In the case of WASP-21, we also considered the three light curves obtained by \citet{barros2011}. These do not cover complete transits, so we converted them to orbital phase before analyzing them in the same manner as described above (see Table \ref{pho-res_w21}).

% Table 5 *** HAT-P-16

\begin{table*}

\centering

\begin{tabular}{lccccc}

\hline\hline

Source & $r_{A}+r_{b}$& $k$ & $i$ & $r_{A}$ &$r_{b}$\\

\hline %

Cassini (transit \#1)   & $0.1416 \pm 0.0040$ & $0.1062 \pm 0.0008$ & $88.45 \pm 0.96$ & $0.1280 \pm 0.0035$ & $0.01359 \pm 0.00043$ \\ %

Cassini (transit \#2)   & $0.1549 \pm 0.0068$ & $0.1085 \pm 0.0009$ & $86.72 \pm 0.85$ & $0.1398 \pm 0.0060$ & $0.01516 \pm 0.00078$ \\ %

CA\,1.23m (transit \#1) & $0.1407 \pm 0.0011$ & $0.1032 \pm 0.0005$ & $89.96 \pm 0.34$ & $0.1275 \pm 0.0010$ & $0.01316 \pm 0.00012$ \\ %

CA\,1.23m (transit \#2) & $0.1485 \pm 0.0054$ & $0.1116 \pm 0.0011$ & $87.28 \pm 0.79$ & $0.1335 \pm 0.0048$ & $0.01491 \pm 0.00065$ \\ %

\hline %

{\bf Final results} & $\mathbf{0.1441 \pm 0.0025}$ & $\mathbf{0.1067 \pm 0.0014}$ & $\mathbf{87.74 \pm 0.59}$ & $\mathbf{0.1303 \pm 0.0022}$ & $\mathbf{0.01377 \pm 0.00038}$ \\ %

\hline

\citet{buchhave2010}       & 0.1542                & $0.1071 \pm 0.0014$ & $86.6 \pm 0.7$ & 0.1392 & 0.0149 \\

\hline

\end{tabular}

\caption{Photometric properties of the HAT-P-16 system derived by fitting the light curves with {\sc jktebop}. The final parameters are given in bold and are compared with those found by \citet{buchhave2010}.}%

\label{pho-res_hp16}%

\end{table*}

%

% Table 6 *** WASP-21

\begin{table*}

\centering

\begin{tabular}{lccccc}

\hline\hline

Source & $r_{A}+r_{b}$& $k$ & $i$ & $r_{A}$ &$r_{b}$\\

\hline %

Cassini                  & $0.1166 \pm 0.0084$ & $0.0998 \pm 0.0016$ & $87.18 \pm 1.00$ & $0.1050 \pm 0.0075$ & $0.01048 \pm 0.00091$ \\

CA\,1.23m                & $0.1182 \pm 0.0057$ & $0.1037 \pm 0.0010$ & $86.83 \pm 0.57$ & $0.1071 \pm 0.0050$ & $0.01110 \pm 0.00062$ \\

\cite{barros2011}        & $0.1166 \pm 0.0042$ & $0.1086 \pm 0.0009$ & $87.01 \pm 0.44$ & $0.1052 \pm 0.0037$ & $0.01142 \pm 0.00049$ \\ %

\hline %

{\bf Final results}      & $\mathbf{0.1169 \pm 0.0031}$ & $\mathbf{0.1055 \pm 0.0023}$ & $\mathbf{86.97 \pm 0.33}$ & $\mathbf{0.1057 \pm 0.0028}$ & $\mathbf{0.01117 \pm 0.00035}$ \\

\hline

\citet{bouchy2010}       & 0.1046              & $0.10820_{-0.00035}^{+0.00037}$ & $88.75_{-0.70}^{+0.84}$ & 0.0948 & 0.00983 \\

\citet{barros2011}       & 0.1149              & $0.10705_{-0.00086}^{+0.00082}$ & $87.34 \pm 0.29 $       & 0.1038 & 0.01112 \\

\citet{southworth2012}   & $0.1186 \pm 0.0042$ & $0.1095 \pm 0.0014$             & $86.77 \pm 0.45 $       & $0.1069 \pm 0.0037$ & $0.01170 \pm 0.00054$ \\

\hline

\end{tabular}

\caption{Photometric properties of the WASP-21 system derived by fitting the light curves with {\sc jktebop}. The light curves from \citet{barros2011} were combined in phase and then analyzed. The final parameters are given in bold and are compared with those found by other authors.}%

\label{pho-res_w21}%

\end{table*}

%%%%%%%%%%%%%%%%%%%%%%%%%%%%%%%%%%%%%%%%%%%%%%%%%%%%%%

\subsection{New orbital ephemerides}%

\label{sec_3.1}

%%%%%%%%%%%%%%%%%%%%%%%%%%%%%%%%%%%%%%%%%%%%%%%%%%%%%%

% Table 4

\begin{table*} %

\centering %

\tiny

\begin{tabular}{lrrl}

% columns

\hline\hline

Time of minimum    & Epoch & Residual & Reference  \\

BJD(TDB)$-2400000$ &       &  (JD)    &            \\

\hline %

$ 55027.59293 \pm 0.00031 $          &   0 &  0.000117 & \cite{buchhave2010} \\

$ 55085.88780 \pm 0.00049  $         &  21 & -0.000409 & \cite{buchhave2010} \\

%$ 55097.0064  \pm 0.0088   $         &  25 &  0.014273 & \cite{buchhave2010} \\

%$ 55124.75507 \pm 0.00763  $         &  35 &  0.003266 & \cite{buchhave2010} \\

$ 55135.853622 \pm 0.000504$         &  39 & -0.002069 & \cite{buchhave2010} \\

$ 55388.46897 \pm  0.00123 $         & 130 & -0.000101 & Trnka  (TRESCA)     \\

$ 55463.42067 \pm  0.00049 $         & 157 & -0.000985 & Vra\v{s}t\'{a}k (TRESCA) \\

$ 55463.4193  \pm  0.0008  $         & 157 &  0.000375 & Vil\'{a}gi, Gajdo\v{s} (TRESCA)  \\

$ 55471.7491  \pm  0.0007  $         & 160 &  0.000902 & Shadick (TRESCA)     \\

$ 55477.30172 \pm  0.00149 $         & 162 &  0.001569 & Vra\v{s}t\'{a}k (TRESCA)   \\

$ 55482.85087 \pm 0.00066  $         & 164 & -0.001223 & Shadick (TRESCA)     \\

$ 55485.6291  \pm 0.0005   $         & 165 &  0.001065 & Sanchez (TRESCA)     \\

$ 55499.50837 \pm 0.00019 $          & 170 &  0.000453 & This work (Loiano 152\,cm) \\

$ 55796.53707 \pm 0.00034 $          & 277 &  0.000230 & This work (Calar Alto 123\,cm)  \\

$ 55829.84931 \pm 0.00059 $          & 289 &  0.000812 & Shadic (TRESCA)      \\

$ 55835.40206 \pm 0.00091 $          & 291 & -0.001041 & Sauer  (TRESCA)      \\

$ 55835.3994 \pm 0.0014   $          & 291 &  0.001619 & Trnka  (TRESCA)      \\

$ 55843.72852 \pm 0.00081 $          & 294 &  0.000165 & Shadic (TRESCA)      \\

$ 55904.79696 \pm 0.00065 $          & 316 & -0.002761 & Shadic (TRESCA)      \\

$ 55935.3302  \pm 0.0012  $          & 327 & -0.005204 & Garcia (TRESCA)      \\

$ 55968.64736 \pm 0.00106 $          & 339 &  0.000301 & Shadic (TRESCA)      \\

$ 56201.83115 \pm 0.00103 $          & 423 &  0.002512 & Shadic (TRESCA)      \\

$ 56204.604209\pm 0.000318 $         & 424 & -0.000404 & This work (Calar Alto 123\,cm)  \\

$ 56204.604513\pm 0.000296 $         & 424 & -0.000100 & This work (Loiano 152\,cm)  \\

%$ 56257.33931 \pm 0.00145 $          & 443 & -0.008489 & This work (Salerno 30\,cm)\\

\hline %

\end{tabular}

\caption{Transit mid-times of HAT-P-16 and their residuals.}\label{res_hp16}%

\end{table*}

% Table 5

\begin{table*}

\centering %

\tiny

\begin{tabular}{lrrl}

% columns

\hline\hline

Time of minimum    & Epoch & Residual & Reference  \\

BJD(TDB)$-2400000$ &       &   (JD)   &            \\

\hline %

$ 54743.0419  \pm 0.0022  $     &   0 &  0.001310  & \cite{bouchy2010}     \\

$ 54743.0283  \pm 0.0062  $     &   0 & -0.012210  & \cite{barros2011}     \\

$ 55084.51951 \pm 0.00032 $     &  79 & -0.000004  & \cite{barros2011}     \\

$ 55438.9709  \pm 0.0011  $     & 161 &  0.004859  & Evans      (TRESCA)   \\

$ 55525.4130  \pm 0.0024  $     & 181 & -0.003433  & \cite{barros2011}     \\

$ 55797.73268 \pm 0.00097 $     & 244 & -0.002409  & Shadic     (TRESCA)   \\

$ 56169.4727  \pm  0.0015 $     & 330 &  0.001008  & Gajdo\v{s} (TRESCA)   \\

$ 56182.43915 \pm 0.00036 $     & 333 & -0.000100  & This work (Calar Alto 123\,cm) \\

$ 56182.43986 \pm 0.00079 $     & 333 &  0.000616  & This work (Loiano 152\,cm) \\

$ 56260.24459 \pm 0.00044 $     & 351 &  0.000002  & Ivanov, Sokov (TRESCA)\\

\hline %

\end{tabular}

\caption{Transit mid-times of WASP-21 and their residuals.}\label{res_w21} %

\end{table*}

During our analysis, we estimated the central transit time of each of our light curves. We enlarged the sample by considering other mid-transit times available in the literature or on websites such as the TRESCA (TRansiting ExoplanetS and CAndidates) archive, which essentially contain light curves obtained by amateur astronomers. We selected only the light curves with a Data Quality index higher than 3 (see Tables \ref{res_hp16} and \ref{res_w21}). Armed with these times of minimum light, we made a linear fit of all the collected mid-transit times as a function of their epoch. We obtained the following ephemeris for HAT-P-16:

% hat-p-16
\begin{equation}
T_{0} = \mathrm{BJD(TDB)}\; 2\,455\,027.59281\,(40) + 2.7759712\,(15)\,E, \nonumber
\end{equation}
and for WASP-21:
%
% wasp-21
\begin{equation}
T_{0} = \mathrm{BJD(TDB)}\; 2\,454\,743.04054\,(71) +4.3225186\,(30)\,E. \nonumber
\end{equation}
The numbers in brackets are the uncertainties to be referred at the last two digits of the number they follow, and $E$ is the number of the cycles after the reference epoch. The quality of the fit is relatively poor ($\chir = 3.68$ and $3.24$), implying that the uncertainties in the individual timings are underestimated. We have increased our quoted uncertainties to reflect this.

\begin{figure*}%
\centering
\includegraphics[width=17.cm]{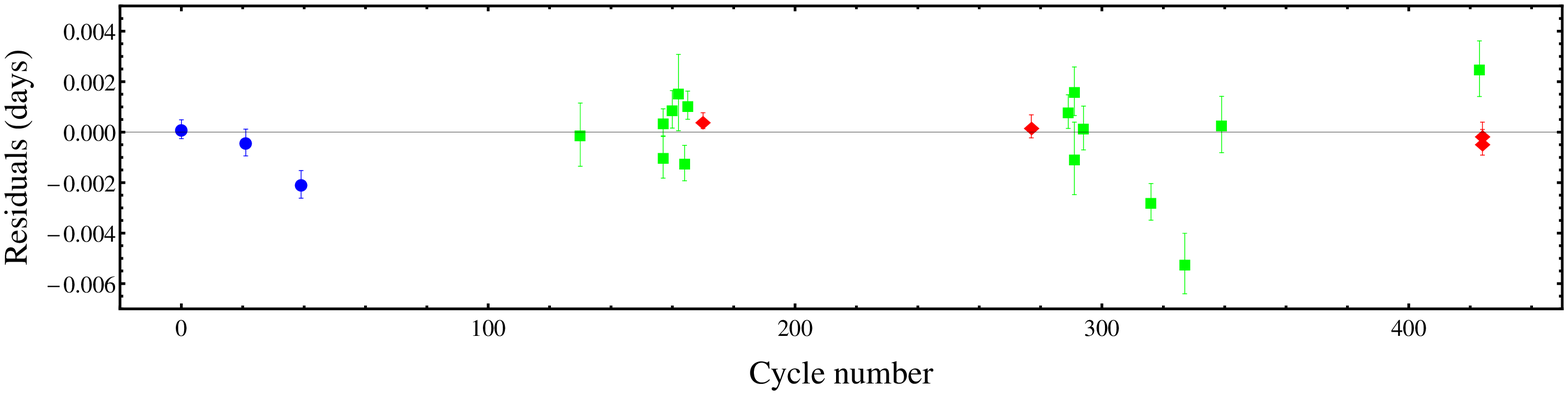}
\caption{Residuals of the timing of mid-transit of HAT-P-16 versus
a linear ephemeris. The different colors blue, green and red
stands for data found in literature, data obtained
from the TRESCA catalog and our data, respectively.}%
\label{oc_hp16}
\end{figure*}

\begin{figure*}%
\centering
\includegraphics[width=17.cm]{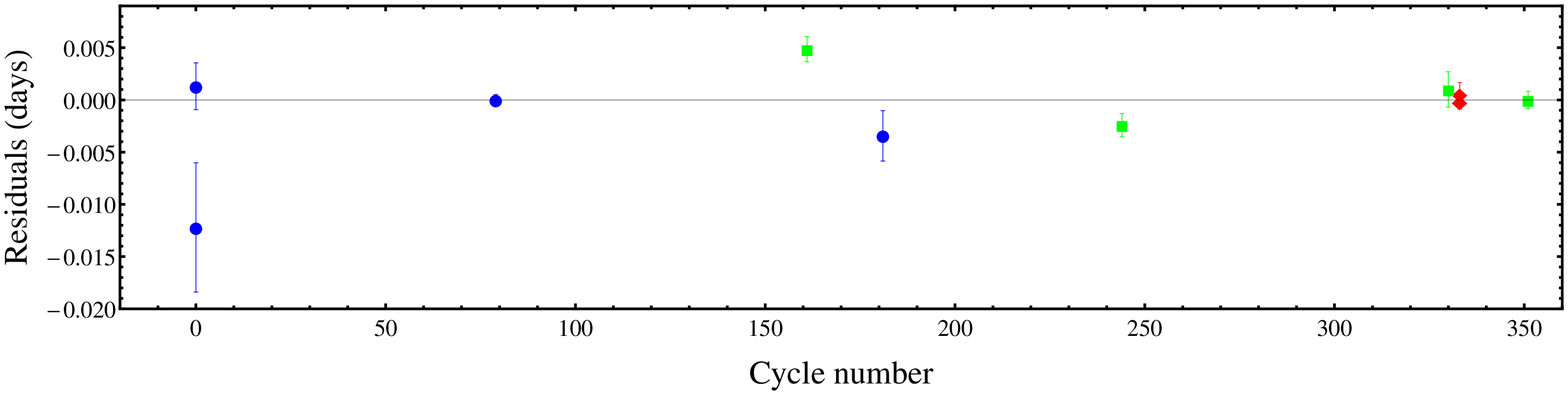}
\caption{Residuals of the timing of mid-transit of WASP-21 versus
a linear ephemeris. The different colors blue, green and red
stands for data found in literature, data obtained
from the TRESCA catalog and our data, respectively.}%
\label{oc_w21}
\end{figure*}

The central times of the transits are also useful to check for the presence of additional bodies. If another planetary object is a member of the system, it should gravitationally interact with the known planet, causing a periodical variation in $T_0$. Other phenomena can cause timing variations, for example starspots \citep{barros2013}. We plot the residuals of the linear fits to the times of minimum light in Figs.\ \ref{oc_hp16} and \ref{oc_w21}. In both cases we do not find any clear evidence of periodic variations in the transit timings.

% Figure 05
%
% Figure 06
%

%%%%%%%%%%%%%%%%%%%%%%%%%%%%%%%%%%%%%%%%%%%%%%%%%%%%%%

\subsection{Final photometric parameters}

\label{sec_3.2}

%%%%%%%%%%%%%%%%%%%%%%%%%%%%%%%%%%%%%%%%%%%%%%%%%%%%%%

Each light curve was modeled separately, yielding its own set of best-fitting parameter values. To assign uncertainties to these values we executed 10,000 Monte Carlo simulations and took the central 68\% of the distribution of simulation parameter values to represent 1$\sigma$ uncertainties. We also calculated uncertainties using a residual-permutation algorithm \citep{southworth2008}, which is sensitive to correlated noise in light curves. We then took the larger of the Monte Carlo or residual-permutation errorbar for each parameter.

The individual results for each light curve of one TEP were then combined into a single set of final photometric parameters. We did this by taking the weighted mean for each parameter, and increasing the errorbar in those cases where the $\chi^2_\nu$ of the weighted mean was greater than unity. The final parameters are given in Tables \ref{pho-res_hp16} and \ref{pho-res_w21}. We found that the agreement between different light curves was good for WASP-21 and acceptable for HAT-P-16. The exception is $k$ for WASP-21, which has $\chi^2_\nu = 14.2$. This value is large, but is not excessive when compared to the results for many of the TEPs analyzed within the Homogeneous Studies project \citep{southworth2012}. A likely source of the discrepancy is spot activity on the host star.

%%%%%%%%%%%%%%%%%%%%%%%%%%%%%%%%%%%%%%%%%%%%%%%%%%%%%%

\section{Physical properties of HAT-P-16 and WASP-21}

\label{sec_4}

%%%%%%%%%%%%%%%%%%%%%%%%%%%%%%%%%%%%%%%%%%%%%%%%%%%%%%

We have measured the main parameters of the two planetary systems, to give a comprehensive picture of their physical characteristics. To perform this step, we used standard theoretical models, the photometric parameters derived in the previous section, and the best spectroscopic parameters available in the literature (summarized in Table \ref{tab:spec}).

\begin{table} %
\centering %
\caption{\label{tab:spec} Spectroscopic properties of the host
stars in HAT-P-16 and WASP-21 adopted from the literature and used
in the determination of the physical properties of the systems.}
\begin{tabular}{lcccc}
\hline
Source      & HAT-P-16 & Ref. & WASP-21 & Ref. \\
\hline
$T_{\mathrm{eff}}$ (K)             & $6140 \pm 72$     & 1 & $5800 \pm 100$   & 3 \\
$[\frac{\mathrm{Fe}}{\mathrm{H}}]$ & $0.12 \pm 0.08$   & 1 & $-0.46 \pm 0.11$ & 3 \\
$K_{\rm A}$ (m\,s$^{-1}$)          & $531.1 \pm 2.8$   & 2 & $116.7 \pm 2.2$  & 3 \\
$e$                                & $0.036 \pm 0.004$ & 2 & 0                & 3 \\ %
\hline %
\end{tabular} %
\tablefoot{(1) \citet{torres2012}; (2) \citet{buchhave2010}; (3)
\citet{bouchy2010}. }
\end{table}
Physical properties were calculated following the approach developed by \citet{southworth2009}. In short, we began with an initial estimate of the stellar mass and iteratively adjusted it to find the best agreement between the calculated stellar radius and observed $T_{\mathrm{eff}}$ versus those predicted by stellar models for this mass. This process was performed for a grid of stellar ages between the zero-age and terminal-age main sequence in order to find the overall best fit.\\
We ran the analysis using each of five different sets of theoretical stellar models (see \citealt{southworth2010}) in order to determine the variation in results arising from stellar theory, and also propagated the errorbars of the input parameters using a perturbation analysis. This yielded a set of physical properties for each system, a model-based age estimate, and separate statistical and systematic errorbars. These results are shown in Tables \ref{fin-res_hp16} and \ref{fin-res_w21}.

% Table 7
\begin{table*}
\centering
\setlength{\tabcolsep}{2pt}
\begin{tabular}{lcc}
\hline\hline

& This work (final) & \!\!\!\!Buchhave et al.\ (2010) \\
\hline %
$M_{\mathrm{A}}$ ($M_{\sun}$)               & $ 1.216 \pm 0.042 \pm 0.036$   & $ 1.218 \pm 0.039$    \\
$R_{\mathrm{A}}$ ($R_{\sun}$)               & $ 1.158 \pm 0.023 \pm 0.011$   & $ 1.237 \pm 0.054$    \\
$\log g_{\mathrm{A}}$ (cgs)                 & $ 4.396 \pm 0.016 \pm 0.004$   & $ 4.34  \pm 0.03 $    \\
$\rho_{\mathrm{A}}$ ($\rho_{\sun}$)         & $ 0.784 \pm 0.040 $            & $ - $                 \\
$M_{\mathrm{b}}$ ($M_{\mathrm{jup}}$)       & $ 4.193 \pm 0.098 \pm 0.083$   & $ 4.193 \pm 0.094$    \\
$R_{\mathrm{b}}$ ($R_{\mathrm{jup}}$)       & $ 1.190 \pm 0.035 \pm 0.012$   & $ 1.289 \pm 0.066$    \\
$g_{\mathrm{b}}$ ($\mathrm{ms^{-2}}$)       & $ 73.4  \pm 4.1   $            & $ 63.1  \pm 5.8  $    \\
$\rho_{\mathrm{b}}$ ($\rho_{\mathrm{jup}}$) & $ 2.33  \pm 0.20  \pm 0.02 $   & $ 1.95  \pm 0.28 $    \\
$T_{\mathrm{eq}}$ ($\mathrm{K}$)            & $ 1567  \pm 22    $            & $ 1626  \pm 40   $    \\
$\Theta$                                    & $ 0.2391\pm0.0073 \pm0.0024$   & $ 0.220 \pm 0.011$    \\
$a$ (AU)                                    & $ 0.04130$ $\pm$ $0.00047$ $\pm$ $0.00041$  & $ 0.0413\pm0.0004$    \\
Age (Gyr)                                   & $0.5 _{-0.5 \,-0.5 }^{+0.4 \,+0.5 }$& $ 2.0   \pm 0.8  $    \\
\hline
\end{tabular}
\caption{Physical properties of the HAT-P-16 system obtained in this work and compared with the discovery paper. The first errorbar given in our results gives the statistical uncertainty and the second refers to the systematic uncertainty.}
\label{fin-res_hp16}%
\end{table*}
%
% Table 8
\begin{table*}
\centering
\begin{tabular}{lcccc}
\hline\hline
& This work (final) & Bouchy et al. 2010 & Barros et al. 2011  & Southworth 2012 \\
\hline %
$M_{\mathrm{A}}$ ($M_{\sun}$)              & $ 0.890 \pm 0.071 \pm 0.035$     & $ 1.01 \pm 0.03$        & $0.86\pm0.04 $            & $ 0.98  \pm 0.12  \pm 0.07$    \\
$R_{\mathrm{A}}$ ($R_{\sun}$)              & $ 1.136 \pm 0.049 \pm 0.015$     & $ 1.06 \pm 0.04$        & $1.097_{-0.022}^{+0.035}$ & $ 1.186 \pm 0.081 \pm 0.028$   \\
$\log g_{\mathrm{A}}$ (cgs)                & $ 4.277 \pm 0.025 \pm 0.006$     & $ 4.39 \pm 0.03$        & $4.29 \pm 0.02$           & $ 4.281 \pm 0.031 \pm 0.010$   \\
$\rho_{\mathrm{A}}$ ($\rho_{\sun}$)        & $ 0.607 \pm 0.048$               & $ 0.84 \pm 0.09$        & $0.65\pm0.05$             & $ 0.587 \pm 0.061$             \\
$M_{\mathrm{b}}$ ($M_{\mathrm{jup}}$)      & $ 0.276 \pm 0.018 \pm 0.007$     & $ 0.300\pm 0.011$       & $0.27\pm0.01 $            & $ 0.295 \pm 0.027 \pm0.014$    \\
$R_{\mathrm{b}}$ ($R_{\mathrm{jup}}$)      & $ 1.162 \pm 0.052 \pm0.015$      & $ 1.07 \pm 0.06$        & $1.143_{-0.030}^{+0.045}$ & $ 1.263 \pm 0.085 \pm0.029$    \\
$g_{\mathrm{b}}$ ($\mathrm{ms^{-2}}$)      & $ 5.07  \pm 0.35$                & $  -  $                 & $5.13 \pm 0.23$           & $ 4.58  \pm 0.45 $             \\
$\rho_{\mathrm{b}}$ ($\rho_{\mathrm{jup}}$)& $ 0.165 \pm 0.018 \pm 0.002$     & $ 0.24 \pm 0.05$        & $0.181_{-0.020}^{+0.015}$ & $ 0.137 \pm 0.021 \pm0.003$    \\
$T_{\mathrm{eq}}$ ($\mathrm{K}$)           & $ 1333  \pm 28$                  & $  -  $                 & $  -  $                   & $ 1340 \pm 32  $               \\
$\Theta$                                   & $ 0.0267\pm 0.0015\pm 0.0004$    & $  -  $                 & $ - $                     & $ 0.0245\pm 0.0019 \pm 0.0006$ \\
$a$ (AU)                                   & $ 0.0499\pm 0.0013\pm 0.0007$    & $0.052_{-0.00044}^{+0.00041}$ & $  0.0494 \pm 0.0009  $             & $ 0.0516\pm 0.0020 \pm 0.0012$ \\
Age (Gyr)                                  &  $  -  $                         & $ 12 \pm 5$             & $12 \pm 2$                & $- $  \\%$9.8_{-5.5\,-11.7}^{+1.4\,+1.1}$
\hline
\end{tabular}
\caption{Physical properties of the WASP-21 system obtained in
this work and compared with those found in literature. The first
error given in our, and Southworth's results is referred to the
statistical errors whereas the second refers to the systematic
uncertainties.}
\label{fin-res_w21}%
\end{table*}

%
%%%%%%%%%%%%%%%%%%%%%%%%%%%%%%%%%%%%%%%%%%%%%%%%%%%%%%
\section{Summary and Conclusions}
\label{sec_5}
%%%%%%%%%%%%%%%%%%%%%%%%%%%%%%%%%%%%%%%%%%%%%%%%%%%%%%
We obtained simultaneous observations of planetary transits in the HAT-P-16 and WASP-21 TEP systems, with the purpose of improving our knowledge of the physical properties of these two systems. The simultaneous observations were performed at two different sites with two medium-class telescopes operating in defocussing mode, achieving observational scatters $\lesssim 1$ mmag per point in four of the six light curves. Our observational strategy was aimed at detecting anomalies in the light curves, which might be attributable to astrophysical phenomena (e.g.\ star spots), but we did not find any clear evidence of these. Small anomalies are detectable by eye in the residuals of the light curve versus the best fits. Each deviation is present in only one light curve, so can safely be attributed to systematic effects arising from the telescope and instrument, or more likely variations in Earth's atmosphere.

We used the new photometric data to revise the ephemerides and physical parameters of the systems. We found the following results:

%

% HAT-P-16

\begin{itemize}

\item [] HAT-P-16

\item [$\bullet$]We obtained improved estimates of the radius of the star and the planet. The value found for the stellar radius is consistent with the one reported in the discovery paper \citep{buchhave2010}, while the planetary radius is smaller by more than 1$\sigma$. The planet has a larger density and surface gravity than previously thought.

\item [$\bullet$] The planet is colder, and the system is less evolved than previously thought.

\item [$\bullet$] Comparing our result obtained for the planetary radius with the theoretical values predicted by \citet{fortney2007} for a planet at $a=0.045$\,AU, we found that the hot Jupiter size is consistent within $1\sigma$ with the $25\, M_{\mathrm{Earth}}$ core model of a H/He planet of nearly $0.3$ Gyr. However since the predictions by \citet{fortney2007} were made for G-type star and HAT-P-16 is an F-type star, we also compared our results with the predictions for $a=0.02$\,AU. In this case the model that best fits our result (within $2 \sigma$) is the prediction for a gaseous planet with a $100\, M_{\mathrm{Earth}}$ core. Both cases suggest that HAT-P-16 is likely a heavy-element rich planet.

\end{itemize}

%

% WASP-21

\begin{itemize}

\item [] WASP-21

% We find a younger age for this planetary system compared to previous studies.

\item [$\bullet$]  We found that the planetary radius is greater than that measured in the discovery paper \citep{bouchy2010}, in agreement with the studies by \citet{barros2011} and \citet{southworth2012}.

\item [$\bullet$]  We compared our result with the prediction made by \citet{fortney2007}, for a similar case: our results are consistent within $3\sigma$ with a core-less model for a H/He-dominated planet. This discrepancy may be due to the unavailability of theoretical predictions of a planet with the exact characteristics of those we measure for WASP-21\,b. Indeed, the lower metallicity of the host star implies that the planet is composed of lighter elements and therefore has a bigger radius. Another possible explanation is proposed by \citet{fortney2007} (and references therein): a planet which was formerly more massive, but has experienced mass loss and became a Neptune-like planet, may have a radius that significantly exceeds $1 R_{\mathrm{jup}}$.

\end{itemize}

\begin{acknowledgements}

Based on observations collected at the Centro Astron\'{o}mico Hispano Alem\'{a}n (CAHA) at Calar Alto, Spain, operated jointly by the Max-Planck Institut f\"{u}r Astronomie and the Instituto de Astrof\'{i}sica de Andaluc\'{i}a (CSIC), and on observations obtained with the 1.52\,m Cassini telescope at the OAB Observatory in Loiano, Italy. The reduced light curves presented in this work will be made available at the CDS (http://cdsweb.u-strasbg.fr/). JS acknowledges financial support from STFC in the form of an Advanced Fellowship. We thank Ulli Thiele and Roberto Gualandi for their technical assistance at the CA 1.23\,m telescope and Cassini telescope, respectively. The following internet-based resources were used in research for this paper: the ESO Digitized Sky Survey; the NASA Astrophysics Data System; the SIMBAD data base operated at CDS, Strasbourg, France; and the arXiv scientific paper preprint service operated by Cornell University. LM thanks the Departamento de Astronom\'ia y Astrof\'isica of the Pontificia Universidad Cat\'olica de Chile for kind hospitality.

\end{acknowledgements}

\bibliographystyle{aa}

\end{document}